# Neural Networks to Solve Partial Differential Equations: A Comparison With Finite Elements

ANDREA SACCHETTI[1], BENJAMIN BACHMANN[2], KASPAR LÖFFEL[2], URS-MARTIN KÜNZI[3], AND BEATRICE PAOLI[3]
[1]IMN, University of Applied Science and Arts Northwestern Switzerland, 8051 Windisch, Switzerland
[2]IPPE, University of Applied Science and Arts Northwestern Switzerland, 8051 Windisch, Switzerland
[3]Laboratory of Web Science, Fernfachhochschule FFHS, 8005 Zurich, Switzerland

Corresponding author: Andrea Sacchetti (andrea.sacchetti@fhnw.ch)

This work was supported by the Hasler Foundation (https://haslerstiftung.ch/).

**ABSTRACT** We compare the Finite Element Method (FEM) simulation of a standard Partial Differential Equation thermal problem of a plate with a hole with a Neural Network (NN) simulation. The largest deviation from the true solution obtained from FEM (0.015 for a solution on the order of unity) is easily achieved with NN too without much tuning of the hyperparameters. Accuracies below 0.01 instead require refinement with an alternative optimizer to reach a similar performance with NN. A rough comparison between the Floating Point Operations values, as a machine-independent quantification of the computational performance, suggests a significant difference between FEM and NN in favour of the former. This also strongly holds for computation time: for an accuracy on the order of $10^{-5}$, FEM and NN require 54 and 1100 seconds, respectively. A detailed analysis of the effect of varying different hyperparameters shows that accuracy and computational time only weakly depend on the major part of them. Accuracies below 0.01 cannot be achieved with the ''adam'' optimizers and it looks as though accuracies below $10^{-5}$ cannot be achieved at all. In conclusion, the present work shows that for the concrete case of solving a steady-state 2D heat equation, the performance of a FEM algorithm is significantly better than the solution via networks.

## I. INTRODUCTION

Partial Differential Equations (PDE) govern the behaviour of most physical systems as shown by prominent examples like the Maxwell, diffusion, Navier-Stokes, elasticity, and heat equations. In most practical applications, analytical solutions are not available and one must resort to numerical methods. These include, but are not limited to, finite-element method (FEM), finite differences, gradient discretisation, finite volume, and method of lines. These standard techniques are employed in many commercial packages for physical simulations with FEM being widely the most commonly used. The use of such packages to predict products' behaviour and optimize designs, among other things, has become very much standard in modern industry. The major part of the mentioned methods exploit some sort of discretisation of the considered domain with a network relating the discretisation points to each other, i.e. a *mesh*. In applications where material motion is involved, i.e. in solid and fluid mechanics, it can be advantageous to employ so-called mesh-free methods. Many such techniques have been developed since the introduction of the smoothed particles hydrodynamics in 1977 [1] but they seem to stay confined in somewhat a niche compared to more widespread tools like FEM.

A somewhat revolutionary mesh-free approach was proposed in 1998 [2] which relies on neural networks (NN). The main idea behind it is to exploit two fundamental properties of NNs:

- NNs satisfy the universal approximation theorem, i.e. any given continuous function $\mathbb{R}^n \to \mathbb{R}^m$ can be approximated to arbitrary accuracy with a single-hidden-layer NN of sufficient width.
- NNs are infintely-differentiable functions whose derivative can be computed analytically by means of automatic differentiation [3].

These two features make NNs an ideal candidate to approximate the solution of a given PDE on a mesh-less geometry. The *ansatz* consists in training the network by means of

The associate editor coordinating the review of this manuscript and approving it for publication was Yongming Li[ID].

  

**TABLE 1.** Comparison of different open-source PDE-NN available packages. Legend - Back-end: T = tensorflow, P = PyTorch; Boundaries: D = Dirichlet, N = von Neumann, C = Custom; Geometry: CSG = Constructive Solid Geometry, B = Basic (simple shapes only), N = Not implemented. Notes: ∗ = support for PyTorch still buggy, + = support for higher dimension coming soon, % = equation types limited.

| name | Reference | back-end | boundaries | dimensions | geometry |
|---|---|---|---|---|---|
| DeepXDE* | [17] | T, P | D, N, C | any | CSG |
| PyDEns | [18] | T | D | any | N |
| NeuroDiffEq+ | [19] | P | D, N | max. 2 | B |
| IDRLnet% | [20] | P | D, N | any | CSG |

minimising a loss function that is given by the PDE itself computed on a discrete set of points from the considered domain. This is made possible by the fact that the derivatives of the NN are available analytically (instead of numerically) and thus can be computed locally. Boundary conditions, both including the function value or its derivatives, can be accommodated as well, as additional explicit penalties in the loss function.

Since this pioneering work [2], the concept has not been pursued much, mostly due to the fact that NNs were a niche discipline until the late 20$^{th}$ century and the required computational power to exploit their capabilities was often not available. The advent of low-cost high-performance hardware as well as the emergence of standardised machine-learning packages including user-friendly NN-libraries (e.g. scikit-learn [4], tensorflow by Google [5], pytorch by Facebook [6]) powered a recent revival of this concept with the appearance of several papers [7]–[16] and also a few open-source projects [17]–[20]. It is worth to mention at this stage that many of these studies have a common denominator in their motivation, namely that the NN-approach to solve PDEs can offer several advantages over established techniques like FEM in specific contexts:

1) While the computational power required to train a NN can be significantly larger than that needed to solve a FEM problem, the need for a re-mesh upon change of the underlying geometry can give the upper hand to the NN approach in those cases where the geometry is not fixed.
2) The training of a NN can be improved iteratively: if the accuracy of the obtained solution is not satisfactory in a given range, it can be improved by means of an incremental training with additional points added in the inaccurate region.
3) Thanks to the universal approximation theorem, NNs are generally known to have a strong generalisation power, i.e. the trained solution can accurately predict the true function on points it never came across during training. The standard procedure of training and testing allows to accurately quantify the generalisation power of the obtained solution.
4) The absence of a mesh shall make the NN-approach more convenient as the dimensionality of the problem increases. This applies in particular for 3D-problems and/or time-dependent ones. Very high-dimensionality problems like the multi-particles Schrödinger equation might particularly profit from this aspect.

It is quite surprising though that none of the aforementioned studies provide a systematic, quantitative comparison between conventional numerical solution of a PDE (e.g., via FEM) versus solution via NN in order to prove empirically the existence of said advantages and pin down the conditions under which they occur. In this study, we try to make a first effort to fill this gap by comparing solutions of the heat equation via FEM and a NN in a 2D domain.

## II. MATERIALS AND METHODS

As a first step, we selected the packages we are using in our study. For FEM, it does not really matter which among the numerous available commercial packages one selects as they are all mature and extremely optimized. The only concern about it, is that this fact might skew the comparison somewhat in favour of FEM because the available packages for PDE-NN are all in their alpha development-stage. This point must be taken into account in the overall comparison. For FEM, we used therefore the standard tool in our team which is the well-established Abaqus package [21]. For the NN-based solution of PDEs we compared four packages that can be found in tab. 1. In order to ensure a fair comparison, all computations have been carried on the same machine (Intel i7-7600U 2.80 GHz, 8.00 GB RAM, Windows 10).

The package "DeepXDE" currently has the clear upper hand in our analysis: it supports any type of equation in unlimited dimensions with arbitrary boundary conditions, it comes with a powerful built-in engine for constructing the underlying geometry, and it has many additional features and options. Moreover, the developers are very active and constantly improving it while a small community is growing around it. The other packages all miss the one or the other of these features and it was decided not to consider them at this stage. Their development shall nevertheless be monitored carefully in case one of them leaps forward significantly or "DeepXDE" comes to a halt. In particular "IDRLnet" looks like a possible contender but it is yet in its early development stages and it still lags behind "DeepXDE" at the moment.

In order to compare the PDE-NN approach with FEM we need a relevant sample problem to solve with both methods. Ideally, this shall be a problem admitting an analytical solution and must possess sufficient complexity to provide a reliable test but at the same time it may not require too large of a computational power so that repeated testing is possible. We identified to this purpose a standard example in introductory FEM theory, namely the stationary heat propagation on a squared two-dimensional plate with a circular hole in the

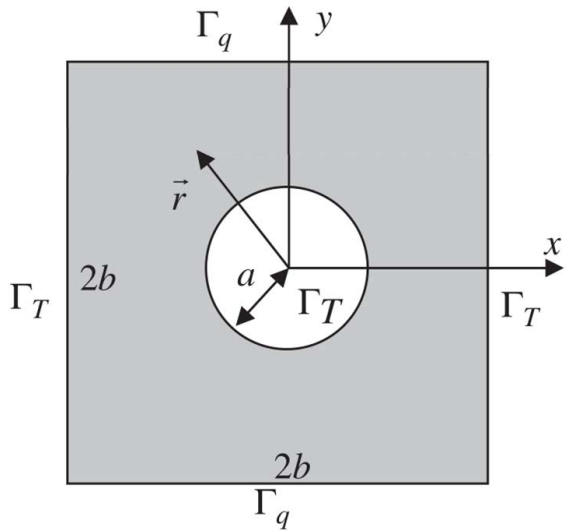

**FIGURE 1.** Domain and boundaries of the heat-propagation problem defined on a squared plate with a hole [22].

centre as in example 8.4 from [22]. The problem reads as follows:

$$\frac{\partial^2 T}{\partial x^2} + \frac{\partial^2 T}{\partial y^2} + \frac{2a}{\sqrt{x^2+y^2}} - 4 = 0$$

The choice of the non-homogenous term ensures that the exact analytical solution is known and equals

$$T(x, y) = \left(\sqrt{x^2+y^2} - a\right)^2.$$

The boundary problem is defined on the squared box of side length $2b$ with a circular hole in the middle of radius $a$. The boundary conditions are of the Dirichlet ($\Gamma_T$) or von Neumann type ($\Gamma_q$) as shown in fig. 1.

The mathematical formulation of the boundary conditions thus reads:

$$T(r = a) = 0$$
$$T(x = \pm b, y) = (\sqrt{b^2 + y^2} - a)^2$$
$$\frac{\partial T}{\partial y}(x, y = \pm b) = \mp 2b \left(\frac{2a}{\sqrt{x^2+b^2}} - 1\right)$$

As it can be noted, both Dirichlet and von Neumann conditions are defined.

## III. RESULTS AND DISCUSSION

As a starting reference point, we take the FEM simulations carried out with Abaqus on the same problem and try to reproduce the performance with NN approach. In the following, we shall always use $a = 0.2$ and $b = 1$.

The FEM simulations have been carried out with five different meshes with varying levels of resolution. In addition, linear and quadratic elements have been used. One example for a relatively coarse mesh (FEM run 1) is represented in fig. 2. The solution in the region of the hole is not computed with FEM and has been padded in the figure. The results of these simulations are summarized in tab. 2. The computational time has been measured in real time, the deviation with respect to the exact solution is root-mean-square averaged on a 50 × 50 grid of points within the domain with points within the hole being excluded from the computation, and the number of Floating Point Operations (FLOPs) is obtained directly from Abaqus. We focus in particular on runs 1 and 10 because these represent the extremal points of the convergence analysis, namely they are the simplest and the most accurate simulations, respectively.

We then proceeded to implement the same problem within the "DeepXDE" package and solve it with the aim of achieving a similar accuracy as with FEM. After several test-runs we found hyperparameter-sets that closely reproduce the accuracy of the aformentioned FEM simulations in run 1. These include a choice of the number of training points comparable with the number of FEM elements in run 1 as well as a similar distribution between boundary and domain. The remaining parameters were optimized for accuracy in training, test, and verification on the grid.

The results of the simulations 1-5 are shown in the tab. 3. The simulation was repeated 5 times to check the reproducibility of the results. It is apparent that the NN-simulations has a similar performance as FEM in terms of accuracy on a grid of points (column "Solution Deviation"). It is also worth to notice that this happens despite the fact that the test error (column "Test error") is systematically larger that the deviation on the grid. This can be traced back to the fact that these two types of errors are indeed not directly comparable: the test error quantifies how much the condition given by the PDE deviates from 0 averaged over the test points and it is in principle not a straightforward measurement of how accurately the NN-solution approximates the true one.

We can also notice a significant discrepancy between the estimated number of FLOPs for the NN-training compared with the FEM-solver, despite of the fact that the computational times are significantly shorter for NN. Part of this discrepancy is possibly due to overhead processes in the FEM calculation under Abaqus as suggested by the fact that computational time stays constant for all but the last two simulation runs with FEM. Moreover the FLOPs value for the NN-simulations is affected by several possible sources of error as it merely estimates the FLOPs for a single model prediction and multiplies it for the estimated number of iterations. This procedure does not take into account for instance any internal optimization (like vectorization) and the estimate for the model, as obtained from Tensorflow, currently lacks a comparison with a theoretical value.[1] It is therefore to be considered as an upper limit.

Reproducing the accuracy of the FEM solution on run 10 turned out particularly challenging with the accessible parameters: the convergence is usually fast but improving the accuracy beyond 0.01 proved to be almost impossible. We thus resorted to accessing hidden optimization parameters and used a different optimization algorithm for refinement: i.e. on top of the "adam" algorithm we performed a subsequent refinement optimization with "L-BFGS-B." (see tab. 4).

---
[1]An empirical dependence of the model-FLOPs for a given layer structure has been established by estimating the FLOPs with tensorflow for different network width and depth.

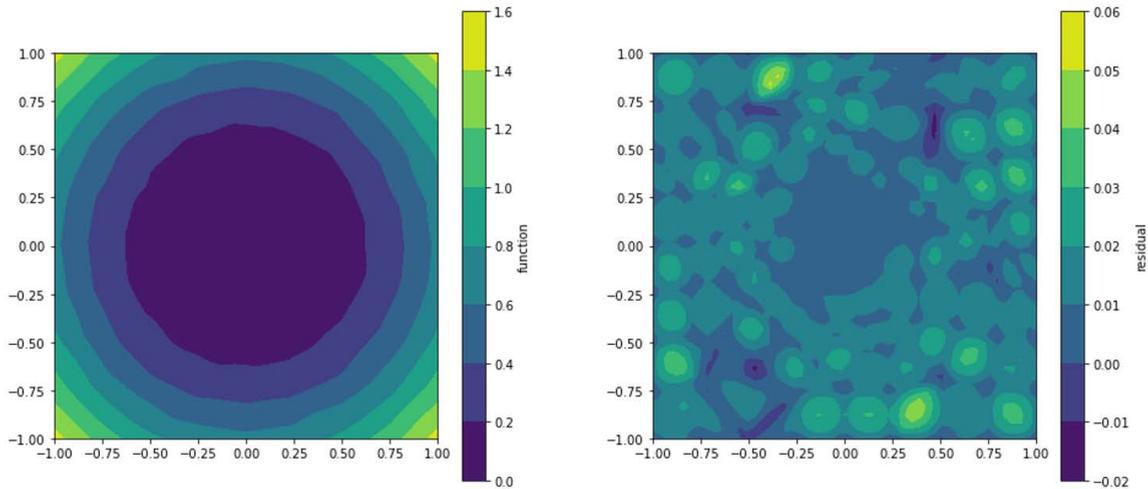

**FIGURE 2.** Left: FEM solution of the heat-propagation problem on a squared plate with a hole for a coarse mesh. Right: Deviation of the FEM solution from the exact one. The region of the hole has been padded in both plots.

**TABLE 2.** Parameters and results of the reference FEM analysis. "Computation time" includes both meshing and solving. FLOPs (floating point operations) only refers to solving.

| Run | Element order | Nodes | Solution deviation | Computation time | FLOPs |
|---|---|---|---|---|---|
| 1 | linear | 115 | 1.476e-02 | 11 s | 2.26e+04 |
| 2 | quadratic | 327 | 1.625e-03 | 11 s | 1.73e+05 |
| 3 | linear | 172 | 9.128e-03 | 11 s | 3.84e+04 |
| 4 | quadratic | 494 | 8.720e-04 | 11 s | 3.86e+05 |
| 5 | linear | 505 | 3.346e-03 | 11 s | 1.85e+05 |
| 6 | quadratic | 1476 | 3.613e-04 | 11 s | 1.72e+06 |
| 7 | linear | 1847 | 1.199e-03 | 11.5 s | 1.33e+06 |
| 8 | quadratic | 5475 | 2.51e-04 | 11.5 s | 1.19e+07 |
| 9 | linear | 46981 | 2.255e-04 | 36 s | 1.65e+08 |
| 10 | quadratic | 141230 | 5.499e-05 | 54 s | 1.41e+09 |

**TABLE 3.** Hyperparameters of the first simulation block (top panel) and corresponding results (bottom panel). "Computational time" consists of the sum of compiling and training time.

| Run | Training pts domain | Training pts boundary | Pts test | Intern layers | Epochs | Optimizer | Learn. rate |
|---|---|---|---|---|---|---|---|
| 1-5 | 50 | 40 | 90 | 20 × 1 | 1000 | adam | 1.00e-02 |

| Run | Test error | Solution deviation | Computational time | FLOPs |
|---|---|---|---|---|
| 1 | 5.13e-02 | 2.01e-02 | 1.9 s | 1.54e+08 |
| 2 | 5.61e-02 | 1.68e-02 | 1.7 s | 1.52e+08 |
| 3 | 4.70e-02 | 2.54e-02 | 2.2 s | 1.50e+08 |
| 4 | 5.54e-02 | 1.72e-02 | 1.5 s | 1.45e+08 |
| 5 | 5.09e-02 | 1.76e-02 | 1.5 s | 1.47e+08 |

In order to reproduce a similar situation upon going from NN-runs 1-5 to 6-10, the number of training points was increased and the network was enlarged, in particular upon increasing the number of layers. Nevertheless, increasing the number of training points towards a similar factor (1300) as in going from FEM runs 1 to 10 turned out to be counterproductive as it significantly affects training time without providing evident benefit in terms of resulting accuracy. It is apparent that the NN-simulations slightly underperforms FEM run 10 in terms of accuracy on a grid of points (column "Solution Deviation"). It is also worth to notice that this time the test error (column "Test error") is smaller than the true-solution deviation on the grid. This can be again traced back to the fact that these two types of errors are indeed not comparable.

The above results call for a more systematic investigation of the effect of the hyperparameters on the simulation's performance. We thus carried out a set of 5500 simulations with random values of the hyperparameters picked within pre-defined ranges in order to explore the effect of single parameters onto the resulting accuracy and computational time.

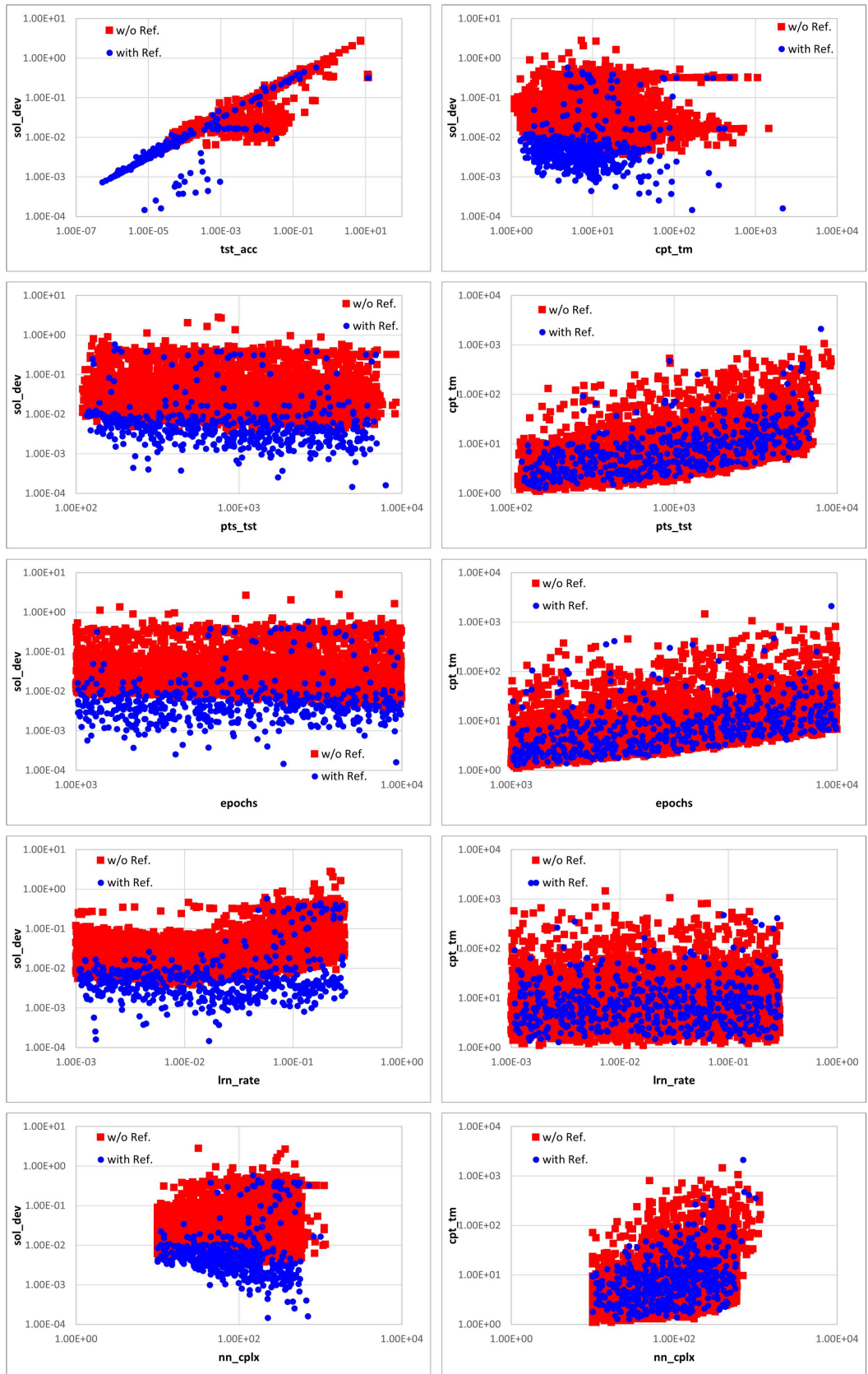

**FIGURE 3.** Summary of the hyperparameters-analysis of computational performance of the NN-method to solve PDEs. Labels legend: sol_dev = rms deviation from true solution, tst_acc = rms accuracy (loss function) on test points, cpt_tm = overall computational time [s], pts_tst = number of test points (equals number of points on domain plus on boundary), lrn_rate = learning rate, nn_cmplx = neural network complexity (depth times width squared). Results with (blue circles) and without (red squares) refinement are shown.

**TABLE 4.** Hyperparameters of the second simulation block (top panel) and corresponding results (bottom panel).

| Run | Training pts domain | Training pts boundary | Pts test | Intern layers | Epochs | Optimizer | Learn. rate |
|---|---|---|---|---|---|---|---|
| 6-10 | 250 | 90 | 340 | $10 \times 4$ | 1000 | adam + L-BFGS-B | 1.00e-02 |

| Run | Test error | Solution deviation | Computational time | FLOPs |
|---|---|---|---|---|
| 6 | 1.42e-04 | 1.99e-03 | 3.8 s | 4.45e+09 |
| 7 | 6.79e-04 | 1.75e-02 | 4.0 s | 4.47e+09 |
| 8 | 1.32e-04 | 1.69e-03 | 3.5 s | 4.38e+09 |
| 9 | 2.42e-04 | 1.60e-03 | 4.3 s | 4.49e+09 |
| 10 | 5.48e-04 | 9.07e-04 | 4.1 s | 4.45e+09 |

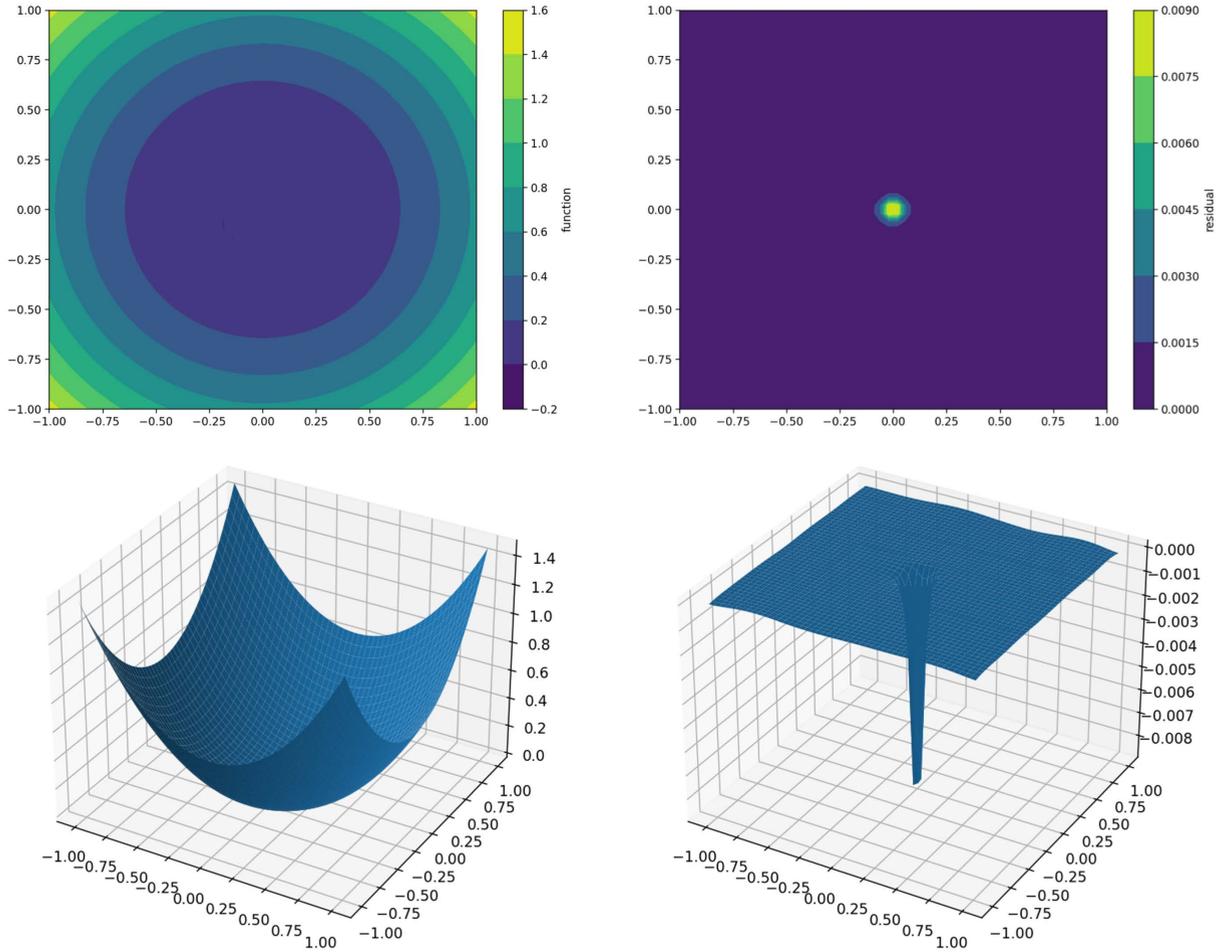

**FIGURE 4.** The best NN-based solution and its residual both as a colour-map and as 3D-plot. The color map of the residual represents its absolute value.

Several indicative plots of the interplay between different hyperparameters and simulation's performance are shown in fig. 3. The main take-away messages from this analysis are the following:

- There is a strong correlation between test accuracy and deviation from the solution but there is a fraction of cases where the solution-deviation is better than what can be predicted from test accuracy.
- Without refinement, accuracy barely correlates with computation time. The correlation with refinement is weak and suggests that a significant improvement in accuracy could require an overproportionally large increase in computation time.
- Similar observations can be made for the number of training points and of epochs: these barely affect accuracy but result in increasing computational time.
- The analysis of the learning-rate dependence suggests that it has a sweet-spot at around 0.03 for the "adam" optimizer. Nevertheless, the subsequent refinement appears to be barely affected by this result.
- Neural network complexity (nn_cplx) was estimated as a product of network depth and square of its width: it seems to have a fair influence on accuracy but it also significantly affects computational time. The exploration of even more complex network will certainly require an increase of the number of training points eventually which implies another computational penalty.

**TABLE 5.** Hyperparameters (top panel) of the direct training to the exact solution and corresponding results (bottom panel).

| run | training pts domain | training pts boundary | pts test | intern layers | epochs | optimizer | learn. rate |
|---|---|---|---|---|---|---|---|
| 1-5 | 50 | 40 | 90 | 20 × 1 | 1000 | adam | 1.00e-02 |
| 6-10 | 250 | 90 | 340 | 10 × 4 | 1000 | adam + L-BFGS-B | 1.00e-02 |

| Run | Test error | Solution Deviation | Computational time |
|---|---|---|---|
| 1 | 4.16e-03 | 6.22e-02 | 0.9 s |
| 2 | 5.22e-03 | 7.65e-02 | 0.9 s |
| 3 | 3.19e-03 | 5.70e-02 | 0.9 s |
| 4 | 4.15e-03 | 5.51e-02 | 0.9 s |
| 5 | 3.98e-03 | 6.28e-02 | 0.9 s |
| 6 | 7.67e-05 | 8.95e-03 | 2.4 s |
| 7 | 5.40e-05 | 7.97e-03 | 2.4 s |
| 8 | 5.05e-05 | 7.04e-03 | 2.9 s |
| 9 | 3.68e-05 | 5.77e-03 | 3.4 s |
| 10 | 2.40e-05 | 5.16e-03 | 3.7 s |

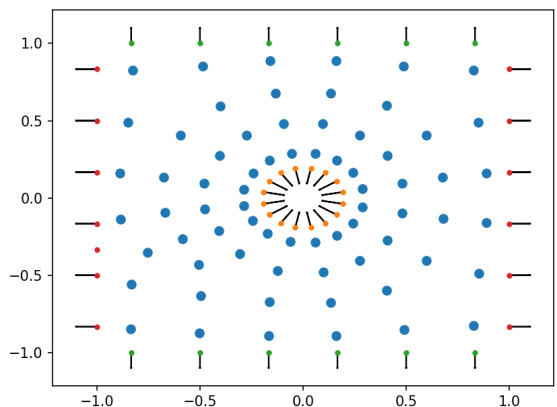

**FIGURE 5.** Centroids of the mesh-elements used as training points for the NN. Blue points: domain; Red, green, and orange points: Boundaries. Boundaries' normal directions for von Neumann conditions are also indicated.

The hyperparameters analysis allowed to determine the optimal set of parameters in terms of accuracy with the minimal computation time. These read as follows - training pts domain: 3600, training pts boundary: 1300, intern layers: 20 × 4, epochs: 10000, learning rate: 0.015, optimizer: "adam" plus refinement with "L-BFGS-B." The simulation carried out with these values returned the record accuracy (solution deviation) of $8.5 \times 10^{-5}$ within a training time of about 1100 s. This deviation is comparable with that of the best FEM solution (run 10). The plots in fig. 4 show the corresponding solution as well as its deviation from the true solution both as a 2D-colour-map and as a 3D-surface. It is apparent that the neural network correctly reproduces the expected radial symmetry of the exact solution and the overall deviation from it has no evident pattern hinting towards the absence of systematic errors. It is important to notice how quickly the result deviates from the analytical solution in the region $r < a$, which is outside the definition domain of the problem. This indicates that extrapolations of the results outside the domain without first extending the training to the regions of interest can be unreliable.

The capability of the neural network was also tested by training it directly against the true solution to assess how well it is capable of reproducing it. Therein, the same parameters as in the previous runs 1-10 have been used. The results are summarized in tab. 5. The chosen neural network can reproduce the true solution to an accuracy that is comparable but worse than that obtained upon solving the PDE, which suggests there is barely any significant potential for improvement in the PDE-based training. This discrepancy can be possibly explained from the fact that the "DeepXDE" package is indeed optimized to solve differential equation and not to fit a known function.

In the following we further cross-check the training on the random points and on the FEM elements. To this purpose, we extract the centroids of the FEM elements and use them as testing points for the NN-model (see fig. 5). As an additional check we train the NN-model on the FEM-centroids and test the result on random-generated points. Both procedures return very similar results thus suggesting that the overall performance of a trained NN-model is globally valid. This also indicates that for more practice oriented problems a hybrid approach can be envisaged where one first solves the base geometry with FEM, thus generating a mesh whose centroids can be used as training points for the NN. Upon exploring changes in the geometry and other physical parameters of the system the NN-solution can be exploited iteratively and the need for a re-mesh, as required by FEM, can be overcome.

## IV. CONCLUSION

In this study we compared quantitatively the accuracy and computational performance of solving a PDE problem by means of FEM and NN for the first time by means of tackling a simple but reasonably realistic problem that admits an analytical solution. Taking the FEM calculation as a reference, the deviation from the true solution can span between 0.015 and $5.5 \times 10^{-5}$. The former value is easily achieved with NN too without much tuning of the hyperparameters. The latter accuracy value resulted unachievable by means of the "adam" optimizer and only a subsequent refinement with the "L-BFGS-B" optimizer allowed to reach a similar accuracy. The reason for this limitation is currently unclear.

While it was not possible to accurately measure the FLOPs required by the NN simulation, thus allowing an absolute comparison of the computational performance, a rough estimate suggests that FEM overperforms NN significantly in this respect, as expected. This is confirmed in particular by the large difference in the computational times for the simulations with the largest accuracy. It is worth to mention once more that the maturity of the available FEM-packages as compared to the PDE-NN ones can explain at least partly why FEM has a clear edge. Moreover, the potential advantages of the NN method are rather expected in subsequent repetitions of the simulation with slight changes of the underlying geometry (like in a design optimisation) than in the single run.

A detailed analysis of the effect of varying different hyperparameters shows that most of them only very weakly affect accuracy and computational time for the NN approach. Accuracies below 0.01 cannot be achieved with the "adam" optimizers and it looks as though accuracies below $10^{-5}$ cannot be achieved at all.

Future work aimed at definitely determining whether this approach is suitable for practical applications shall include the following steps:

- Refine the FLOPs calculation in the NN both theoretically and numerically to ensure a solid and reliable comparison with FEM.
- Compare the two methods on more complex calculations, e.g. upon using a much finer mesh in FEM and targeting a similar accuracy with NN.
- Optionally generate problems with analytical solutions on a more complex geometry: this is possible if one starts from an analytical function and constructs an *ad-hoc* PDE from it, including boundary conditions.
- Find possible ways to introduce a position dependent parameter (i.e. like a material property).
- Elaborate strategies for representative sampling of points from an arbitrary, complex geometry (higher point-density in smaller/sharper regions as well as at interfaces).
- Solve and compare a "real world" examples with particular attention to solid-mechanics problems that represent the vast majority of FEM applications in the practice.

## REFERENCES

[1] R. A. Gingold and J. J. Monaghan, "Smoothed particle hydrodynamics: Theory and application to non-spherical stars," *Monthly Notices Roy. Astron. Soc.*, vol. 181, no. 3, pp. 375–389, Dec. 1977.

[2] R. Š. Hilscher and V. Zeidan, "Hamilton–Jacobi theory over time scales and applications to linear-quadratic problems," *Nonlinear Anal., Theory, Methods Appl.*, vol. 75, no. 2, pp. 932–950, Jan. 2012.

[3] A. G. Baydin, B. A. Pearlmutter, A. A. Radul, and J. M. Siskind, "Automatic differentiation in machine learning: A survey," *J. Mach. Learn. Res.*, vol. 18, pp. 1–43, Apr. 2018.

[4] F. Pedregosa, G. Varoquaux, A. Gramfort, V. Michel, B. Thirion, O. Grisel, and M. Blondel, "Scikit-learn: Machine learning in Python," *J. Mach. Lear. Res.*, vol. 12, no. 85, pp. 2825–2830, Oct. 2011.

[5] M. Abadi. (2015). *Tensorflow.org*. [Online]. Available: https://www.tensorflow.org/

[6] A. Paszke, S. Gross, F. Massa, A. Lerer, J. Bradbury, G. Chanan, and T. Killeen, "PyTorch: An imperative style, high-performance deep learning library," in *Proc. NIPS*, 2019, pp. 1–12.

[7] K. S. McFall and J. R. Mahan, "Artificial neural network method for solution of boundary value problems with exact satisfaction of arbitrary boundary conditions," *IEEE Trans. Neural Netw.*, vol. 20, no. 8, pp. 1221–1233, Aug. 2009.

[8] E. Samaniego, C. Anitescu, S. Goswami, V. M. Nguyen-Thanh, H. Guo, K. Hamdia, X. Zhuang, and T. Rabczuk, "An energy approach to the solution of partial differential equations in computational mechanics via machine learning: Concepts, implementation and applications," *Comput. Methods Appl. Mech. Eng.*, vol. 362, Apr. 2020, Art. no. 112790.

[9] J. Han, E. JentzenWeinan, and A. Jentzen, "Solving high-dimensional partial differential equations using deep learning," *Proc. Nat. Acad. Sci. USA*, vol. 115, no. 34, pp. 8505–8510, 2018.

[10] J. Berg and K. Nyström, "A unified deep artificial neural network approach to partial differential equations in complex geometries," *Neurocomputing*, vol. 317, pp. 28–41, Nov. 2018.

[11] M. A. Nabian and H. Meidani, "A deep learning solution approach for high-dimensional random differential equations," *Probabilistic Eng. Mech.*, vol. 57, pp. 14–25, Jul. 2019.

[12] J. Sirignano and K. Spiliopoulos, "DGM: A deep learning algorithm for solving partial differential equations," *J. Comput. Phys.*, vol. 375, pp. 1339–1364, Dec. 2018.

[13] S. Dong and Z. Li, "Local extreme learning machines and domain decomposition for solving linear and nonlinear partial differential equations," *Comput. Methods Appl. Mech. Eng.*, vol. 387, Dec. 2021, Art. no. 114129.

[14] A. Dedner and R. Klöfkorn, "Extendible and efficient Python framework for solving evolution equations with stabilized discontinuous Galerkin methods," *Commun. Appl. Math. Comput.*, pp. 1–40, Sep. 2021, doi: 10.1007/s42967-021-00134-5.

[15] Y. Shin, "On the convergence of physics informed neural networks for linear second-order elliptic and parabolic type PDEs," *Commun. Comput. Phys.*, vol. 28, no. 5, pp. 2042–2074, Jun. 2020.

[16] D. W. Abueidda, Q. Lu, and S. Koric, "Meshless physics-informed deep learning method for three-dimensional solid mechanics," *Int. J. Numer. Methods Eng.*, vol. 122, no. 23, pp. 7182–7201, Dec. 2021, doi: 10.1002/nme.6828.

[17] L. Lu, X. Meng, Z. Mao, and G. E. Karniadakis, "DeepXDE: A deep learning library for solving differential equations," *SIAM Rev.*, vol. 63, no. 1, pp. 208–228, 2021.

[18] R. Khudorozhkov, S. Tsimfer, and A. Koryagin, "PyDEns framework for solving differential equations with deep learning," 2019. [Online]. Available: https://github.com/analysiscenter/pydens

[19] F. Chen, D. Sondak, P. Protopapas, M. Mattheakis, S. Liu, D. Agarwal, and M. Di Giovanni, "NeuroDiffEq: A Python package for solving differential equations with neural networks," *J. Open Source Softw.*, vol. 5, no. 46, p. 1931, Feb. 2020.

[20] W. Peng, J. Zhang, W. Zhou, X. Zhao, W. Yao, and X. Chen, "IDRLnet: A physics-informed neural network library," 2021, *arXiv:2107.04320*.

[21] *ABAQUS/Standard User's Manual*, Dassault Systèmes Simulia Corp., Johnston, RI, USA, 2021.

[22] J. Fish and T. Belytschko, "Finite element formulation for multidimensional scalar field problems," in *A First Course in Finite Elements*, 1st ed. Chichester, U.K.: Wiley, 2007, ch. 8, sec. 8.2, pp. 205–206.

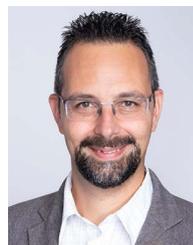

**ANDREA SACCHETTI** was born in Italy, in 1977. He received the B.S., M.S., and Ph.D. degrees in physics from the University of Rome "La Sapienza," in 2001 and 2005, respectively.

From 2005 to 2008, he was a Postdoctoral Fellow at the ETH Zurich. From 2008 to 2018, he collected industrial experience in various research and development-strong companies, such as Bruker Biospin, Sensirion, and Kistler Instruments. Since 2018, he has been a Full Professor of physics at the University of Applied Sciences and Arts Northwestern Switzerland. He authored numerous publications and patents in different fields of basic and applied physics and is also a referee of several international journals. His research interests include applications of experimental physics and of algorithms and simulations to product development and industrial research.

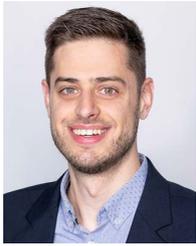

**BENJAMIN BACHMANN** was born in Switzerland, in 1988. He received the B.S. degree in mechanical engineering from the University of Applied Sciences and Arts Northwestern Switzerland (UAS NW), Windisch, Switzerland, in 2013.

Since 2013, he has been working as a Research Assistant at the UAS NW's Institute of Product and Production Engineering and in between gained four years of industry experience as a CAD Expert at Tribecraft. His research interests include FEM simulations, 3D printing, and mechanical design of prototypes for tests and measurements.

Mr. Bachmann is a member of the National Agency for Finite Element Methods and Standards (NAFEMS, U.K.) through the institute.

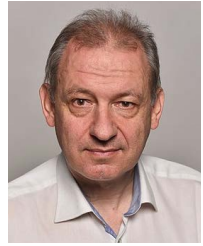

**URS-MARTIN KÜNZI** was born in Zürich, in 1957. He received the M.S. and Ph.D. degrees in mathematics from the University of Zurich, in 1980 and 1984, respectively.

From 1984 to 1986, and in 1988, he was a Postdoctoral Fellow at the University of Bonn. In 1987, he was a Research Fellow at the Academy of Science, Novosibirsk. He was a Postdoctoral Fellow at the University of Freiburg, from 1989 to 1991, and at the University of Berne, from 1992 to 1996. From 1998 to 2007, he worked as a Software Developer. From 2007 to 2014, he was the Leader of the degree program in computer science at the Swiss Distance University of Applied Sciences (FFHS). Since 2015, he has been a Researcher of data science at the FFHS. He authored numerous publications in algebra (quadratic forms), model theory, computability theory, and graph theory. His research interests include machine learning, graph theory, mathematical logic, and software development.

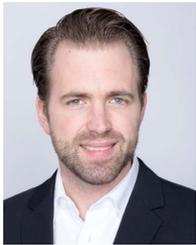

**KASPAR LÖFFEL** was born in Switzerland, in 1982. He received the B.S. and M.S. degrees in structural engineering from ETH Zurich, in 2007, and the Ph.D. degree in mechanical engineering from the Massachusetts Institute of Technology, in 2012.

From 2012 to 2015, he was a Project Engineer at Alstom Switzerland Ltd. Since 2015, he has been a Full Professor of products and production engineering at the University of Applied Sciences and Arts Northwestern Switzerland. His research interests include continuum mechanics of solids, product development, and metal additive manufacturing. He has authored numerous peer-reviewed publications and patent applications in the above-mentioned fields.

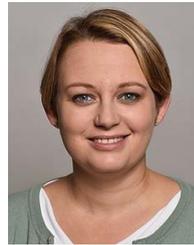

**BEATRICE PAOLI** was born in Italy, in 1981. She received the B.S. and M.S. in physics from the University of Rome ''La Sapienza,'' in 2005, and the Ph.D. degree in computational biochemistry from the University of Zurich, in 2009.

Since 2010, she has been employed at the Swiss Distance University of Applied Sciences (FFHS) first as a Research Fellow and then as a Group Leader and a Full Professor. She leads several projects in the field of data science with applications in different industrial sectors. She authored publications in different research fields, including applied physics, biochemistry simulations, and data science. Her research interests include machine learning, deep learning, and natural language processing.

∙ ∙ ∙